\documentclass[
 aip,
 amsmath,amssymb,
reprint
]{revtex4-1}

\usepackage{graphicx}
\usepackage{dcolumn}
\usepackage{bm}

\usepackage[utf8]{inputenc}
\usepackage[T1]{fontenc}
\usepackage{mathptmx}
\usepackage{etoolbox}
\usepackage{siunitx}
\usepackage{hyperref}

\makeatletter
\def\@email#1#2{
 \endgroup
 \patchcmd{\titleblock@produce}
  {\frontmatter@RRAPformat}
  {\frontmatter@RRAPformat{\produce@RRAP{*#1\href{mailto:#2}{#2}}}\frontmatter@RRAPformat}
  {}{}
}
\makeatother
\begin{document}

\preprint{AIP/123-QED}

\title{Photonic Negative Differential Thermal Conductance Enabled by NIS Junctions}
\author{Matteo Pioldi}
    \email{matteo.pioldi@sns.it}
\author{Giorgio De Simoni}
\affiliation{ 
NEST, Istituto Nanoscienze-CNR and Scuola Normale Superiore, I-56127 Pisa, Italy
}
\author{Alessandro Braggio}
\affiliation{ 
NEST, Istituto Nanoscienze-CNR and Scuola Normale Superiore, I-56127 Pisa, Italy
}
\affiliation{Institute for Quantum Studies, Chapman University, Orange, CA 92866, USA
}

\author{Francesco Giazotto}
\affiliation{ 
NEST, Istituto Nanoscienze-CNR and Scuola Normale Superiore, I-56127 Pisa, Italy
}

\begin{abstract}
Owing to their sensitivity to temperature fluctuations, normal metal-insulator-superconductor (NIS) junctions are leveraged in various thermal devices. This study illustrates that two NISIN reservoirs can achieve a measurable negative differential thermal conductance (NDTC). This phenomenon is enabled by photon-mediated heat exchange, which is profoundly affected by the temperature-dependent impedance matching between the reservoirs. Under appropriate configurations, the heat current is suppressed for increasingly large temperature gradients, leading to NDTC. We also propose experimental configurations where it is possible to discriminate this effect unambiguously. We employ superconducting aluminum in conjunction with either silver or epitaxial InAs to facilitate the experimental observation of NDTC at low temperatures over significant sub-Kelvin ranges. This advances the development of devices that exploit NDTC to enhance heat and temperature regulation in cryogenic environments, such as thermal switches, transistors, and amplifiers.
\end{abstract}

\maketitle

Normal metal-insulator-superconductor (NIS) junctions capitalize on the gapped quasiparticle spectra inherent in superconductors, enabling a myriad of applications \cite{pekola2024nisreview}. At sufficiently low temperatures, they can be harnessed for single-electron charge pumps~\cite{pekola2008pumptheory, pekola2008pumpexp, pekola2009pumphigh, kemppinen2009pumptheory, zorin2009pumpdissipative, pekola2022pumppower}, thermometers~\cite{tsui1976thermofirst, pekola2006mesoscopics,zumbuhl2015thermomillik, pekola2015thermofast, pekola2015thermoandreev, pekola2020thermosnis}, bolometers~\cite{golubev2001bolometer, kalabukhov2003bolometer, kuzmin2020bolometer}, as well as refrigerators for electronic~\cite{martinis1994refrignis, averin1996refrigsinis,quaranta2011cooling, pekola2012refrigreview, pekola2016refrigcascaded} and photonic~\cite{mottonen2017phrefrig, mottonen2024singlenisphrefrig} degrees of freedom. Furthermore, they serve as tunable components in circuit quantum electrodynamics~\cite{mottonen2018phemission, mottonen2018phlamb, jerome2022phhighimp, mottonen2024qubitreset} and in qubit reset protocols~\cite{mottonen2022qubitreset, tsai2023qubitreset}. Notably, the charge transport properties of NIS junctions exhibit significant temperature sensitivity, which underpins state-of-the-art thermometry~\cite{pekola2023thermo}.
\begin{figure}
\includegraphics[width=3.37in]{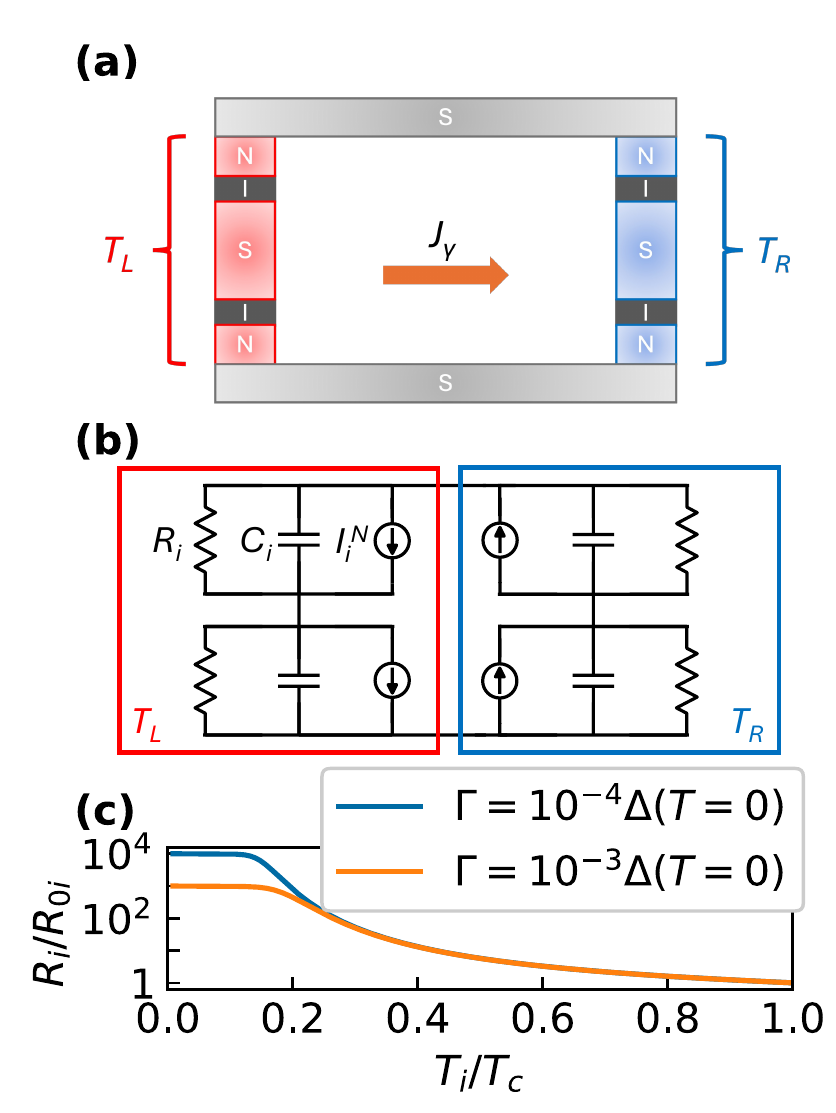}
\caption{\label{fig:1}(a) A diagram of the proposed setup features two NISIN junctions, with the normal, insulating, and superconducting components denoted by the letters N, I, and S, respectively. The "Left" junction is in red and maintained at temperature $T_L$, while the "Right" junction is blue and set at temperature $T_R$. The superconducting lines are shown in grey. The arrow indicates the direction in which the heat flow $J_{\gamma}$ is positive. (b) The lumped element circuit diagram of the setup shown in (a). (c) $R_i(T_i)/R_{0i}$ as a function of temperature for distinct values of the Dynes parameter $\Gamma$.}
\end{figure}

Temperature-sensitive impedances can lead to observing unconventional behaviors in heat transport mediated by electromagnetic fluctuations (a "photonic heat channel"). In this context, heat exchange is enabled by the spontaneous noise radiation from resistive reservoirs connected through dissipationless superconducting lines \cite{cleland2004radiative, pascal2011radiative}. Photonic heat channels are significant at low temperatures~\cite{giazotto2021diode}and over long distances~\cite{pekola2006radiativeexp, mottonen2016radiativedistance}. In this framework, the effectiveness of heat exchange heavily relies on matching the impedances between the involved thermal reservoirs. If this matching is temperature-dependent, configurations may arise where the heat flow between two reservoirs increases despite a decrease in their temperature difference, a phenomenon known as "negative differential thermal conductance" (NDTC) (or "anomalous heat conduction"~\cite{su2025mpemba}). NDTC can be leveraged to create devices that control heat currents and temperatures, \textit{e.g.} thermal oscillation amplifiers \cite{casati2006ndtc, giazotto2016amplifier, giazotto2025photonic}. 

In this Letter, we propose to use a photonic heat channel between two NISIN junctions, harnessing their temperature-dependent impedances, to realize NDTC in a realistic cryogenic solid-state device. This method is similar to Ref. \onlinecite{marchegiani2023ndtc}, which exploited the superconducting phase transition. However, our proposal extends NDTC to broader and more tunable ranges of temperatures, which are relevant for applications. Additionally, we outline arrangements to promptly achieve the experimental demonstration of the discussed NDTC at cryogenic temperatures.

In the proposed setup, two NISIN junctions (the "Left" and "Right" reservoirs) are connected by two lossless superconducting lines, as illustrated in Fig. \ref{fig:1}(a). 
The superconducting lines (in grey) suppress heat transport through quasiparticle diffusion \cite{cleland2004radiative} because their clean contacts with normal metal operate ideally as perfect Andreev mirrors \cite{Andreev1964} at subgap energies. At sub-Kelvin electronic temperatures, electron thermalization with the substrate phonon bath is significantly reduced \cite{giazotto2017caloritronicsreview}, and the photonic heat channel becomes progressively relevant in the heat exchange between reservoirs. 
NISIN junctions allow fabrication using only one superconductor. Instead, SINIS reservoirs require lines made of a material with a gap larger than that of the superconducting part of the reservoir to prevent the diffusion of hot quasiparticles from one side to the other. Anyway, the framework we will outline can be adapted to configurations where the reservoirs are SINIS or NIS junctions.

We consider a linear approximation for the impedance across NIS junctions 
\cite{feldman1985linearNIS, mottonen2022linearnis}, self-consistently verifying that the voltage fluctuations are sufficiently small for this assumption to hold. Each NIS junction consists of a temperature-dependent tunnel resistance $R_i(T_i)$, a capacitor $C_i$, and a generator of noise current $I_i^N(T_i)$, where $i=L,R$ (the two junctions of each reservoir are assumed to be identical). We neglect the intrinsic resistance of the normal components and spurious reactances along the superconducting lines. Fig. \ref{fig:1}(b) illustrates the simplified lumped element scheme of the entire circuit. The linearized $R_i(T_i)$ is obtained by computing the quasiparticle current for a small voltage $\delta V$ using the standard "semiconductor" model for the NIS~\cite{tinkham2004}~\footnote{We assume to neglect any environmental influence on the tunneling if not eventually included in the phenomenological Dynes' parameter\cite{pekola2006mesoscopics}}:
\begin{eqnarray}
    \delta I(T_i)&=&\frac{1}{2 e R_{0i}}\int_{-\infty}^{\infty}\!\!\! \! \!  d\epsilon N_S(\varepsilon, T_i) [f_i(\varepsilon - e\delta V) - f_i(\varepsilon + e\delta V)]  \nonumber\\
    &\approx&\frac{\delta V}{R_i(T_i)},
\end{eqnarray}
where $f_i(\varepsilon)=1/(1+\exp[\varepsilon/(k_B T_i)])$, and $R_{0i}$ is the normal-state tunnel resistance of the junction. The normalized superconducting density of states \cite{dynes1978dos} is
$
    N_S(\varepsilon, T_i) =| \operatorname{Re} [ (\varepsilon - i \Gamma)/\sqrt{(\varepsilon - i \Gamma)^2 - \Delta(T_i)^2} ]|,
$
with $\Gamma$ the superconductor's Dynes parameter. The BCS temperature-dependent superconducting gap is~\cite{tinkham2004}
$
    \Delta(T_i) \approx 1.764 k_B T_c \tanh (1.74 \sqrt{T_c/T_i - 1}),
$
with $T_c$ the critical temperature of the superconductor. 

The ratio $R_i(T_i)/R_{0i}$, plotted in Fig. \ref{fig:1}(c) as a function of $T_i/T _c$ for two different values of $\Gamma$, is a BCS universal function, \textit{i.e.} independent of the specific superconductor. As shown in the plots, $R_i(T_i)/R_{0i}$ depends on $\Gamma$ only for $T \lesssim 0.2T_c$. We take $\Gamma = 10^{-4} \Delta(T=0)$, but our results are quite robust to variations of $\Gamma$ since we typically operate at $T > 0.2 T_c$.

We utilize the framework from Ref. \onlinecite{pascal2011radiative} to calculate $J_{\gamma}$, the heat current flowing from the left to the right reservoir. We assess it as the sum of the photonic heat balances of the two junctions on the left, yielding:
\begin{equation}
    J_{\gamma}(T_L,T_R) = \int_0^{\infty} \frac{\hbar\omega}{2 \pi} \tau_{LR}(\omega) [n_L(\omega) - n_R(\omega)] \text{d}\omega,
\end{equation}
where $n_i(\omega) = 1/(\exp[\hbar \omega / (k_B T_i)] - 1)$ and $\tau_{LR}$ is the transmission coefficient
    \begin{equation}
    \tau_{LR}(\omega) = \frac{4 R_L(T_L) R_R(T_R)}{[R_L(T_L) + R_R(T_R)]^2 + [\omega R_L(T_L) R_R(T_R) (C_L + C_R)]^2}.
    \label{eq:transmission}
\end{equation}
These formulae evaluate the noise emitted by the reservoirs using the fluctuation-dissipation theorem, which is a valid model for NIS junctions in thermal equilibrium \cite{levitov1996noise, scalapino1974noise, buttiker2000noise, safi2016noise} and in the linear regime in voltages.
In such a limit, the heat current $J_{\gamma}$ depends not on the number of junctions in each reservoir but rather on their total impedance. Furthermore,  $J_{\gamma}$ is also scale-invariant with the junction areas $A_i$ since $R_i$ ($C_i$) we take resistances (capacities) to be inversely (directly) proportional to the areas of the corresponding junctions. We can define the specific resistances $r(T_i) = R_i(T_i) A_i$ (capacitances $c = C_i /A_i$). Finally, the transmission coefficient (see Eq.~\ref{eq:transmission}) can be rewritten in terms of the area mismatch $\chi = A_R / A_L$, $r(T_i)$ and $c$:
\begin{equation}
    \tau_{LR}(\omega) = \frac{4\chi r(T_L) r(T_R)}{[\chi r(T_L) + r(T_R)]^2 + [\omega/\omega_{c}]^2},
\end{equation} 
wherein we introduced a cutoff frequency $\omega_{c}=[(1+\chi)c\sqrt{r(T_L) r(T_R)}]^{-1}$.
Interestingly, the $\chi$ factor could also represent a difference in the tunnel junction-specific resistances between the two reservoirs, grasping different physical mechanisms behind the impedance mismatch. 

\begin{figure}
\includegraphics[width=3.37in]{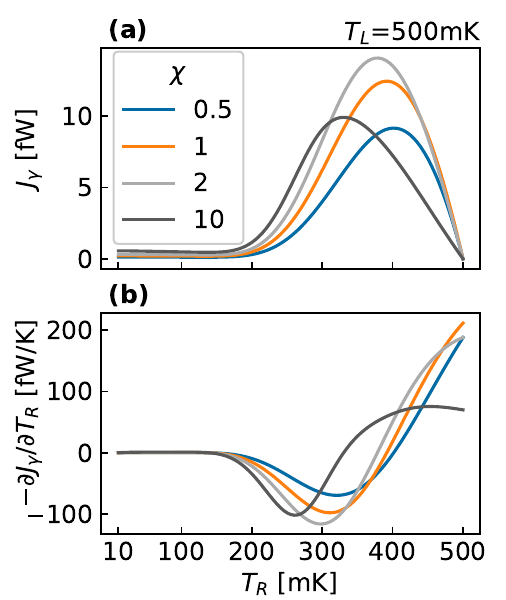}
\caption{\label{fig:2}(a) The heat current traveling from the left to the right terminal of the photonic channel $J_{\gamma}$, with $T_L=\qty{500}{\milli \kelvin}$ fixed as $T_R$ is varied, for different values of $\chi$. (b) The differential thermal conductance $-\partial J_{\gamma}/\partial T_R$, under the same conditions as in (a).}
\end{figure}

In Fig. \ref{fig:2}(a) $J_{\gamma}$ is plotted keeping fixed $T_L=\qty{500}{\milli \kelvin}$ while varying $T_R$ between $\qty{10}{\milli \kelvin}$ and $T_L$, for  different values of $\chi$. For the remainder of this Letter, we will consider $\qty{10}{\milli \kelvin}$, a typical substrate temperature for cryogenic experiments, as the lower limit of the temperature ranges.
It is important to note that the right reservoir at $T_R$ in this configuration takes the role of the cold terminal, so if it becomes hotter, one would generally expect a reduction in the heat current. Instead, an increase in the current is a clear signature of NDTC. 
\begin{figure*}
\includegraphics[width=6.74in]{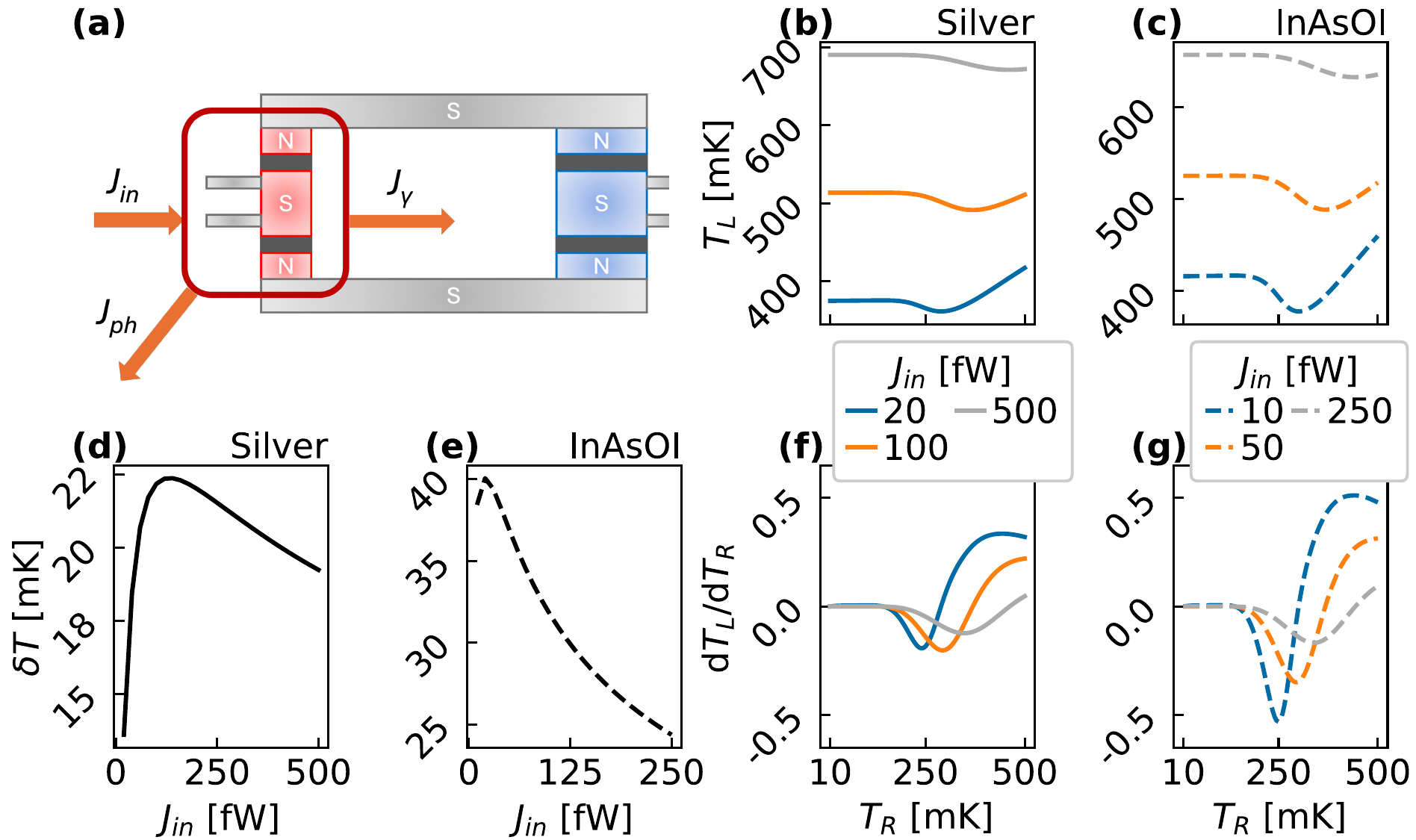}
\caption{\label{fig:3} (a) The first setup, where $T_R$ is varied using heaters (small grey contacts on the right). Arrows represent the direction in which the corresponding heat currents are taken positive. 
For each quantity of interest, plots for both the silver- (solid lines) and the InAsOI-based (dashed lines) setups are presented, with values of $J^L_{in}$ chosen to achieve similar $T_L$s. 
(b)-(c) $T_L$ as a function of $T_R$ for different $J^L_{in}$s. (d)-(e) The difference $\delta T$ between $T_L(T_R = \qty{10}{\milli \kelvin})$ and $T_{L,min}$ for different $J_{in}^L$s. (f)-(g) The responsivity $\text{d}T_L/\text{d}T_R$ of the temperature of the left reservoir with respect to variations in the right one. We assume the left reservoir to be made of two silver wires, each measuring $V_{Ag} = 1.3 \times 0.2 \times 0.01 \unit{(\micro \metre)^3}$, and a middle aluminum region of $V_S = 2 \times 0.2 \times 0.01 \unit{(\micro \metre)^3}$. For the InAsOI platform, we employ two InAs mesas, each of $V_{InAs} = 1.3 \times 0.2 \times 0.1 \unit{(\micro \metre)^3}$, with a middle aluminum wire measuring $V_S = 2 \times 0.2 \times 0.02 \unit{(\micro \metre)^3}$. The tunnel junctions on the left are of $A_L = \qty{0.1}{(\micro \metre)^2}$.}
\end{figure*}
Indeed, the non-monotonic behavior of $J_{\gamma}$ as $T_R$ increases can be attributed to two competing effects. On one hand, the transmission coefficient $\tau$ transitions from zero to finite values owing to improved matching between $R_L(T_L)$ and $R_R(T_R)$, which accounts for the initial rise in heat flux. Conversely, elevated $T_R$s result in reduced thermal gradients across the channel. Consequently, this diminishes $J_{\gamma}$ once the transmission coefficient reaches saturation (at frequencies of interest, which need to be lower than both $\omega_{c}$ and $k_B T_L / \hbar$) until $T_L=T_R$ is reached, at which point heat current necessarily ceases entirely.
The graphs illustrate that $\chi$ significantly influences the performance of the photonic transport channel. In circumstances where $\chi \gg 1$, $\omega_{c}$ is reduced due to the increasing mismatch of the imaginary components of the total junction impedances and, in such a case, $\tau \sim 1/\chi$. Conversely, when $\chi \ll 1$, $\tau \sim \chi$ occurs.
Furthermore, in instances where $\chi < 1$, there is a more significant mismatch in resistances for $T_R \le T_L$, further decreasing $J_{\gamma}$ but shifting toward higher temperatures the maximum heat current. On the other hand, in the case of $\chi \gtrsim 1$, the two sides achieve an alignment at lower values of $T_R$, increasing the maximum heat current. This shows that the nonlinearity of the problem implies that perfectly identical junctions are not necessarily optimal, and a difference in their areas could be helpful. Indeed, higher values of $J_{\gamma}$ are attainable due to efficient matching under increased thermal gradients. The optimal value is determined to be $\chi \approx 2$.

\begin{figure*}
\includegraphics[width=6.74in]{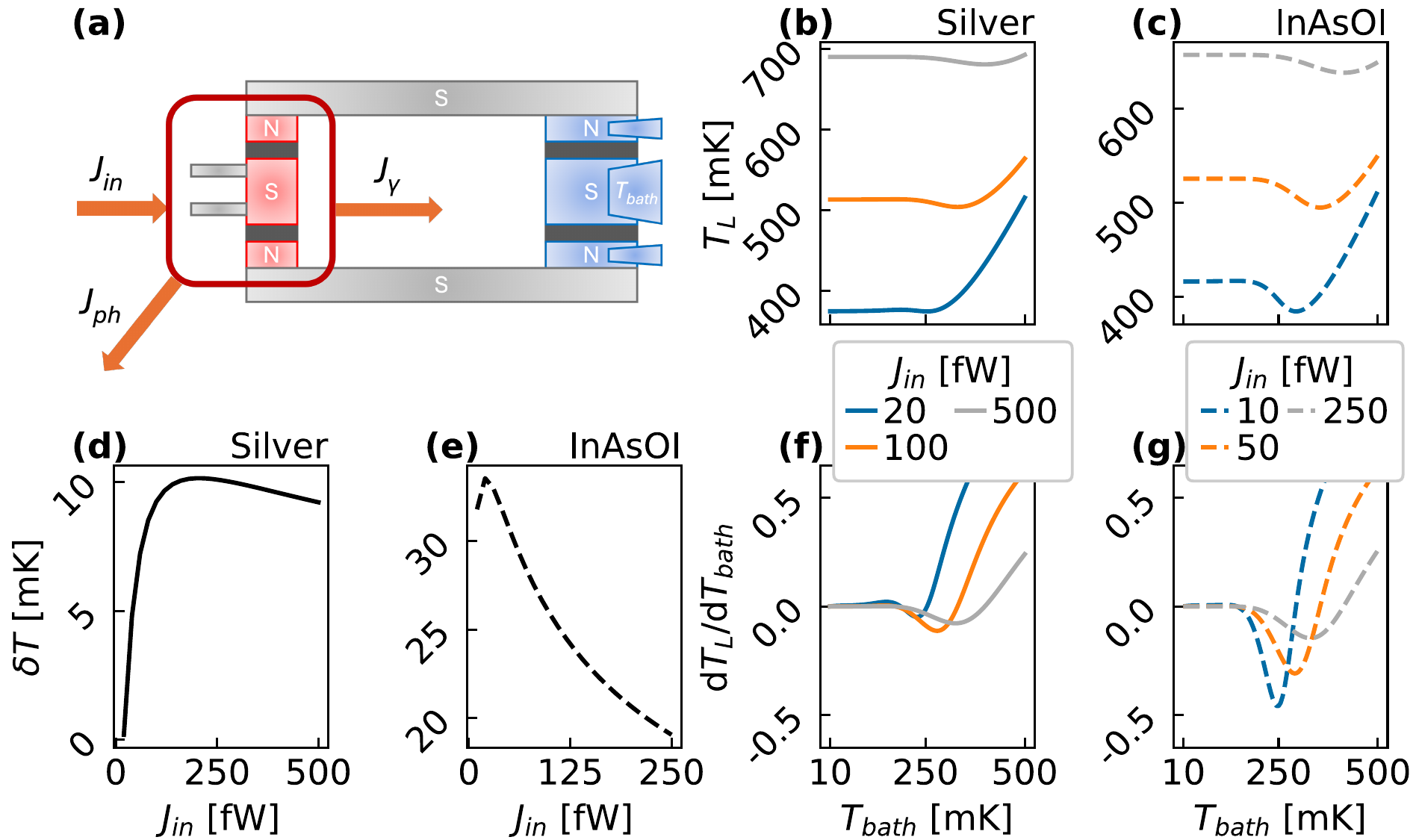}
\caption{\label{fig:4} (a) The second setup, where $T_R = T_{bath}$. Arrows represent the direction in which the corresponding heat currents are taken positive. For each quantity of interest, plots for both the silver- (solid lines) and the InAsOI-based (dashed lines) setups are presented, with values of $J^L_{in}$ selected to have similar $T_L$s. (b)-(c) $T_L$ as a function of $T_{bath}$ for different $J^L_{in}$s. (d)-(e) The difference $\delta T$ between $T_L(T_{bath} = \qty{10}{\milli \kelvin})$ and $T_{L,min}$ for different $J_{in}^L$s. (f)-(g) The responsivity $\text{d}T_L/\text{d}T_{bath}$ of the temperature of the left reservoir with respect to variations of the bath. Sizings are the same as Fig. \ref{fig:3}.}
\end{figure*}

By computing \cite{casati2006ndtc} $\kappa_{\gamma} = -\partial J_{\gamma}/\partial T_R$ and plotting it in Fig. \ref{fig:2}(b), we observe that this heat channel exhibits NDTC. For each curve corresponding to a different value of $\chi$, the interval of $T_R$ where NDTC is observed aligns with the range where the corresponding flux in Fig. \ref{fig:2}(a) increases. The most negative values occur for $\chi \approx 2$. 

In the following, we discuss two experimental setups designed to provide clear evidence of NDTC\cite{marchegiani2023ndtc}. Specifically, we propose observing how the temperature $T_L$ of the left reservoir, which receives a fixed input power, decreases as the right reservoir becomes hotter. This is only possible if, as NDTC suggests, the heat flow mediated by the photonic channel increases despite the reduction of the thermal gradient at its terminals, which effectively reduces the electronic temperature of the hot terminal. This effect cannot be explained without NDTC.

To evaluate $T_L$, we use the steady-state balance of heat currents at thermal equilibrium, assuming negligible Kapitza resistance:
\begin{equation}
    J_{in}^L - J_{\gamma}(T_L,T_R) - \sum_k J_{ph,k}^L(T_L,T_{bath}) = 0,
\end{equation}
where $J_{in}^L$ is the fixed input power provided to the left NISIN (which takes the role of the hot terminal) and $\sum_k J_{ph,k}^L$ represents the total heat exchange of the electrons from the components ($k=N,S$) of the left reservoir with the substrate phonon bath at temperature $T_{bath}$. 
The capability of photon-mediated heat transport to operate over sizable distances \cite{mottonen2016radiativedistance} enables spatial separation of the NISINs, thus ruling out heat exchange mediated by the substrate itself.
To maximize $J_{\gamma}$, we consider setups where $\chi=2$.

We propose two distinct material platforms for the NIS junctions: Al/AlOx/Ag and Al/AlOx/InAsOI. InAsOI ("InAs on insulator") is a doped InAs layer epitaxially grown on a metamorphic, cryogenically insulating InAlAs substrate \cite{battisti2024inasoi, paghi2025inasoi, senesi2025inasoi}. We selected these platforms due to their low electron-phonon couplings, as demonstrated in the following. 
The heat exchange with the phonon bath depends on the material, the volume $V$, and the temperatures.
For the normal parts, $J_{ph,N}^L(T_L,T_{bath}) = \Sigma_N V (T_L^n - T_{bath}^n)$, where $n=5$, $\Sigma_{Ag} = \qty{5e8}{\watt \per (\metre^3 \kelvin^5)}$ for silver wires~\cite{pekola2006mesoscopics} and $n=6$, $\Sigma_{InAs} =  \qty{3e7}{\watt \per (\metre^3 \kelvin^6)}$ for InAsOI~\cite{battisti2024inasoi}.
Instead, for superconducting aluminum we take~\cite{timofeev2006phonon} $J_{ph,S}^L(T_L,T_{bath}) =\Sigma_S V_S / (96 \zeta(5) k_B^5) \int_{-\infty}^{\infty} \text{d}\varepsilon \varepsilon N_S(\varepsilon, T_L) \int_{-\infty}^{\infty} \text{d}\varepsilon' \operatorname{sgn}(\varepsilon') \varepsilon'^2 [ \mathcal{F}_{\varepsilon} \cdot \mathcal{F}_{\varepsilon + \varepsilon'} - 1- (\mathcal{F}_{\varepsilon} - \mathcal{F}_{\varepsilon + \varepsilon'})\coth(\varepsilon/(2 k_B T_{bath}))] N_S(\varepsilon + \varepsilon', T_L) (1-\Delta(T_L)^2/(\varepsilon^2 + \varepsilon \varepsilon'))$, with $\zeta$ the Riemann function, $\mathcal{F}_E=\tanh[E/(2 k_B T_L)]$ and $\Sigma_S = \qty{3e8}{\watt \per (\metre^3 \kelvin^5)}$. We will only consider configurations below the critical temperature $T_{c,Al}=\qty{1.2}{\kelvin}$. 
The wires are sufficiently elongated to render proximitization effects negligible. We assume realistic estimates for specific resistance $R_{0i}/A_i=\qty{10}{\ohm (\micro \metre)^2}$ and capacitance $c=\qty{50}{\pico \farad \per (\micro \metre)^2}$, determining impedances sufficiently significant to neglect the intrinsic resistances of the normal wires, the complex impedance of the superconductors \cite{bardeen1958} and charging effects. The first setup is illustrated in Fig. \ref{fig:3}(a). In this scenario, we assume it is feasible by modulating power injection in the right superconductor to tune $T_R$ (cold side) assuming fixed $T_{bath} = \qty{10}{\milli \kelvin}$. At the same time, the power $J_{in}^L$ introduced in the left superconductor (hot side) is assumed to be fixed when $T_R$ is varied.
Thus, heaters are applied on the S sections of the NISIN in both reservoirs in the figure. We verified that heat exchange between the superconductor and normal metal terminals is strong enough for the N and S parts to have the same temperatures $T_i$.  

In Fig. \ref{fig:3}(b)-(c), we present three typical $T_L$ responses to variations in $T_R$ for different values of $J_{in}^L$, as observed in both the silver and the InAsOI devices. Given that the input power $J_{in}^L$ remains constant, the decrease in $T_L$ for $T_R > \qty{200}{\milli \kelvin}$ must be attributed to an increase in the heat flow leaving the left reservoir. Since the power dissipated in the phononic system $\sum_k J_{ph,k}^L$ is unaffected by the electronic temperature $T_R$, the reason for this phenomenon is a more significant $J_{\gamma}$ despite a reduced thermal gradient, \textit{i.e.} NDTC. 

In Fig. \ref{fig:3}(d)-(e) the difference $\delta T = T_L(T_R=\qty{10}{\milli \kelvin}) - T_{L,min}$, where $T_{L,min}$ is the minimum value measured while varying $T_R$ but keeping fixed input power, is plotted as a function of different $J_{in}^L$s. We can identify a broad range of $J_{in}^L$s for each device where a significant variation $\delta T$ can be observed. Due to its minimal coupling with the substrate phonons, InAsOI shows more substantial $\delta T$s. 
Under the $T_L(T_R)$ plots, in Fig. \ref{fig:3}(f)-(g), we also present the responsivity $\text{d}T_L/\text{d}T_R$, which emphasizes the unusual trend in $T_L$ as $T_R$ increases, computed for different values of the input power in the hot terminal $J_{in}^L$. 
These figures also enable us to compare the strength of the thermal response induced by the photonic heat channel at various working points. 
In particular, an input current around $\qty{100}{\femto \watt}$ produces a steeper response in silver. 
This optimal working point is achieved for $T_L \approx \qty{500}{\milli \kelvin}$, where photonic heat transport reaches significant values while the coupling to the phononic bath remains low. 
For higher $T_L$s, the coupling to the phonon bath suppresses the influence of NDTC.

The second setup is illustrated in Fig. \ref{fig:4}(a). In this case, $T_R$ is not independently varied but equals $T_{bath}$. This can be achieved by equipping the right reservoir with prominent cooling fins, as shown in the image.
The specific measurements indicated in Fig. \ref{fig:3}(b)-(g) are analyzed and represented graphically for this configuration in Fig. \ref{fig:4}(b)-(g). In this scenario, introducing heat to the phonon bath promotes a shift in the equilibrium also of the left reservoir toward more elevated temperatures, countering the NDTC, which instead determines a $T_L$ reduction given that the $J_{in}^L$ is kept constant. Also, in this configuration, a reduction of $T_L$ with increasing $T_R$ implies NDTC.
This effect is noticeable on the right side of Fig. \ref{fig:4}(b)-(c), where, following the temperature minimum, $T_L$ increases more rapidly than in the previous setup. Heating the phonon bath decreases $\delta T$. This is evident in Fig. \ref{fig:4}(d)-(e), where $\delta T$ is reduced compared to the previous configuration, especially for the silver-based device.
The absolute values of the negative responsivity decrease, as shown in Fig. \ref{fig:4}(f)-(g). Due to its minimal phonon coupling, these changes are less significant for InAsOI. 

In summary, we proposed a setup where two NISIN reservoirs are used to achieve NDTC in the framework of photon-mediated heat exchange in a robust way with respect to scale and geometry.  NDTC is always present whenever photonic thermal transport occurs between two impedances that tend to align as their temperatures converge.  Then, we outlined experimental protocols to prove NDTC clearly, relating it to otherwise unexplainable temperature reductions with increasing power inflow. We predict the observation of sizable NDTC-related $T_L$ variations of at least $\qty{10}{\milli \kelvin}$ for $J_{in} > \qty{10}{\femto \watt}$, overcoming spurious parasitic power inputs \cite{giazotto2012interferometer}. 
While our proposal allows for significant experimental observation of NDTC at cryogenic temperatures, applications are feasible across various configurations and devices.
For example, it is possible to analyze the performance of such a setup as a thermal diode when the reservoirs are not identical: $\chi \neq 1$, employing superconductors with dissimilar gaps, or substituting one of the NISIN with a NININ (see also Ref. \onlinecite{giazotto2021diode}). 
Furthermore, photonic heat transport is a versatile framework in which tunable impedances can be integrated into the superconducting lines to enable more precise control of heat fluxes~\cite{giazotto2017tunable, pekola2018tunable, pekola2020tunable, giazotto2021tunable, jerome2022phhighimp}.

The potential to achieve NDTC and significantly reduce heat flux between two reservoirs presents various applications. NDTC facilitates the design of thermal transistors, amplifiers, and memories~\cite{casati2006ndtc, jauho2008transistor, giazotto2016amplifier, benabdallah2014thermaltransistor, xie2024thermaldevices, giazotto2025photonic}, and the exploration of out-of-equilibrium effects \cite{pierattelli2024deltatnoise, su2025mpemba}. 
These devices can be utilized for fully photonic thermal logic, as already proposed with phononic 
\cite{li2007phononlogic} and electronic 
\cite{giazotto2018phaselogic} degrees of freedom. 
Furthermore, the control achieved on $J_{\gamma}$ enables switches and regulators \cite{rojo2020thermaldevices} useful to adjust the interaction between autonomous thermal machines (\textit{e.g.} refrigerators) and the hot environment \cite{mitchison2019autonomous, gasparinetti2025reset}.

\begin{acknowledgments}
We acknowledge the EU’s Horizon 2020 Research and Innovation Framework Programme under Grants No. 964398 (SUPERGATE), No. 101057977 (SPECTRUM), and the PNRR MUR project PE0000023-NQSTI for partial financial support. A.B. acknowledges the 
MUR-PRIN 2022—Grant No. 2022B9P8LN-(PE3)-Project NEThEQS “Non-equilibrium coherent thermal effects in quantum systems” in PNRR Mission 4-Component 2-Investment 1.1 “Fondo per il Programma Nazionale di Ricerca e Progetti di Rilevante Interesse Nazionale (PRIN)” funded by the European Union-Next Generation EU and CNR project QTHERMONANO. A. B. thanks Prof. A. N. Jordan, Dr. B. Bhandhari, and A. N. Singh for fruitful discussions. 
\end{acknowledgments}

\section*{Authors Declarations}

\subsection*{Conflict of interest}
The authors have no conflicts to disclose.

\subsection*{Authors contributions}
\textbf{Matteo Pioldi}: Conceptualization (equal); Data curation (lead); Formal analysis (lead); Investigation (lead); Methodology (equal); Software (lead); Visualization (lead); Writing – original draft (lead); Writing – review \& editing (lead). \textbf{Giorgio De Simoni}: Investigation (equal); Supervision (equal); Writing – review \& editing (equal). \textbf{Alessandro Braggio}: Investigation (equal); Formal analysis (equal); Funding acquisition (equal); Resources (equal); Supervision (equal); Writing – review \& editing (equal). \textbf{Francesco Giazotto}: Conceptualization (equal); Formal analysis (equal); Funding acquisition (equal); Investigation (equal); Resources (equal); Supervision (lead); Methodology (equal); Project Administration (lead); Resources (equal); Supervision (equal); Writing – review \& editing (equal).

\section*{Data Availability Statement}
The data supporting this study's findings are available from the corresponding author upon reasonable request.

\end{document}